**Switchable geometric frustration in an artificial-spin-ice/superconductor hetero-system**

Yong-Lei Wang[1,2,3,+,*], Xiaoyu Ma[2,+], Jing Xu[1,4,+], Zhi-Li Xiao[1,4,*], Alexey Snezhko[1], Ralu Divan[5], Leonidas E. Ocola[5], John E. Pearson[1], Boldizsar Janko[2,*], Wai-Kwong Kwok[1]

[1]*Materials Science Division, Argonne National Laboratory, Argonne, Illinois 60439, USA*

[2]*Department of Physics, University of Notre Dame, Notre Dame, Indiana 46556, USA*

[3]*School of Electronic Science and Engineering, Nanjing University, Nanjing 210093, China*

[4]*Department of Physics, Northern Illinois University, DeKalb, Illinois 60115, USA*

[5]*Center for Nanoscale Materials, Argonne National Laboratory, Argonne, Illinois 60439, USA,*

+ Authors contribute equally

* Correspondence to: ylwang@anl.gov; xiao@anl.gov; bjanko@nd.edu

1. **Geometric frustration emerges when local interaction energies in an ordered lattice structure cannot be simultaneously minimized, resulting in a large number of degenerate states. The numerous degenerate configurations may lead to practical applications in microelectronics [1], such as data storage, memory and logic [2]. However, it is difficult to achieve extensive degeneracy, especially in a two-dimensional system [3,4]. Here, we showcase in-situ controllable geometric frustration with massive degeneracy in a two-dimensional flux quantum system. We create this in a superconducting thin film placed underneath a reconfigurable artificial-spin-ice structure [5]. The tunable magnetic charges in the artificial-spin-ice strongly interact with the flux quanta in the superconductor, enabling the switching between frustrated and crystallized flux quanta states. The different states have measurable effects on the superconducting critical current profile, which can be** **reconfigured by precise selection of the spin ice magnetic state through application of an external magnetic field. We demonstrate the applicability of these effects by realizing a reprogrammable flux quanta diode. The tailoring of the energy landscape of interacting 'particles' using artificial-spin-ices provides a new paradigm for the design of geometric frustration, which allows us to control new functionalities in other material systems, such as magnetic skyrmions [6], electrons/holes in two-dimensional materials [7,8] and topological insulators [9], as well as colloids in soft materials [10-13].**

2. Flux-quanta (FQ) in a type II superconductor are quantized magnetic flux bundles, each containing a single flux-quantum, sustained by circulating superconducting currents. Their motion determines the entire electromagnetic behavior of a superconductor [14]. Controlling the distribution and motion of FQ is important for enhancing the superconductor's current carrying capacity and developing innovative superconducting electronic devices [15]. Recent research demonstrates that the FQ system can be a unique platform to investigate geometric frustration [16-18], with certain clear advantages over other widely investigated systems such as artificial-spin-ices (ASI) of nanoscale bar magnets [2,3,5,9,19-26], colloidal-ices of microscopic magnetic particles [10,11], and buckled nonmagnetic colloidal monolayers [12]. In the ice-analogue of FQ system, a pair of closely placed pinning potentials, such as nanopatterned holes, in a superconducting film provides two possible locations for a FQ, mimicking the binary degrees of freedom of a 'spin' in an ASI system [16-18]. Both theoretical [16] and experimental [17,18] studies have shown that strong interactions between FQs could drive the system into the ground state more readily. However, similarly to the two-dimensional (2D) ASI [21-26], the ground state configurations of the FQ ice systems are all characterized by a two-fold degenerate long range ordering [16-18], and thus far, an extensive degenerate ground state has not been realized in such systems.

3. Extensive degeneracy in the ASI system comprised of square lattices of nanoscale bar magnets has been achieved recently by vertically separating two sublattices of orthogonally oriented bar magnets [3], thereby creating a 3D structure. To realize a massively degenerate geometric frustration in a 2D FQ system, we impose a magnetically charged landscape onto a superconductor. This landscape is derived from a recently introduced tri-axial ASI structure consisting of three sublattices of nanomagnetic bars in a square lattice (Extended Data Fig. 1a) [5] and is very effective in controlling the position of FQs in a superconducting thin film (Fig. 1a). The tri-axial ASI produces three types (Type-I, II and III in Figs. 1c-1e) of magnetic charge (MC) order out of eight possible charge configurations (Extended Data Fig. 1b-1i). The MC configurations can be conveniently obtained by applying an in-plane magnetic field (see Method). The ground state Type-I MC order consists of two sublattices of plaquettes (M and N in Fig. 1b), one rotated by $90^{o}$ from the other. This is similar to the arrangement of the orthogonally oriented bar magnets in the 3D ASI [3], but in a planar 2D structure. Depending on the orientation of the magnetic field (i.e. pointing in or out of the superconducting film), a MC can act as an attractive or repulsive potential barrier for the FQs (Fig. 1a) [27]. The Type-I MCs in each plaquette (Fig.1b) produce an array of ice-like 'two attractive, two repulsive' potential center for the FQ, and enables the realization of a true 2D FQ system that possesses geometric frustration with extensive degeneracy (see discussions below).

4. The FQs in a superconducting film can be driven into motion by an electric current (Fig. 1a), enabling dynamic studies of geometric frustration through magneto-transport experiments [17,18]. In Figs. 1f-1h, we show color coded maps of the magnetic field and current dependent dissipation voltage induced by the motion of FQs. Extended Data Fig. 2 shows maps associated with all eight possible MC configurations. The superconducting critical current (black-dashed line), defined as the maximum current a superconductor can carry without energy dissipation, exhibits clear kinks/peaks at certain magnetic fields indicated by the vertical white lines in Fig. 1f-1h. These are typical features of field matching effects [17,18] originating from the collective energy minimization when the FQs form a crystallized lattice commensurate with an ordered lattice of pinning potentials. Here, the FQ density $n$ is determined by the magnetic field $B$ through the relationship $n=B/\Phi_0$, where $\Phi_0=2.07\times10^{-15}$ Wb is the magnetic flux carried by a single FQ. The first matching field is $B/B^\Phi = 1$, where $B^\Phi=40$ Oe corresponds to a FQ density of one FQ per plaquette for our sample. Note that while kinks/peaks are observed at all integer and in some case, half-integer matching fields under the Type-II and III charge configurations, the matching effect is missing at $B/B^\Phi = 1$ under Type-I charge configuration, indicating a disordered arrangement of FQs.

5. To obtain further insight into the FQs matching effect and to determine the distribution of FQs, we carried out extensive molecular dynamic (MD) simulations (see Methods). Since the observed FQ matching effect is determined by MC patterns (see Extended Data Fig. 3), we can simplify our magnetic/superconducting system with a model based on MCs interacting directly with FQs. The simulated critical current curves (Extended Data Fig. 4) reproduce all the matching features shown in our experiments. Furthermore, the simulated FQ distributions (Fig.2 and Extended Data Fig. 5) confirm that the FQs crystallize into ordered states at all the magnetic fields where the peaks/kinks are present, while they show no long range order under Type-I charge configuration at $B/B^\Phi = 1$. Figure 2 also shows that all the FQs sit at the positions of attractive potential centers (blue circles) under Type-I charge configurations. For an arbitrary pair of attractive potential centers, we denote the energetically favorable FQ filling configuration 'one filled, one empty' with a two-fold degeneracy as a 'favorable pair' in Fig. 3a, while the 'two empty' configuration with higher FQ/MC interaction energy and the 'two filled' configuration with higher inter-FQs interaction energy as a 'frustrated pair'. Our simulation for $B/B^{\Phi}=1$ indicates that the favorable pairs of nearest-neighbor attractive centers (α pairs defined in Fig. 3b) within each plaquette are systematically ubiquitous with population up to 80%. When favorable α pairs form, there is no way to simultaneously satisfy all the β pairs in the favorable configuration (Fig. 3b), resulting in geometric frustration between the nearest-neighbor M-N plaquettes and a two-fold degeneracy within each plaquette.

6. In addition to the frustrated local (or short-range) interactions, the breakdown of long range order is required to obtain extensive degeneracy in a geometrically frustrated system [20]. In Figs. 3c-3e, we compared the populations of favorable and frustrated pairwise configurations with those of the noninteracting FQs obtained from the field dependent population rate formula listed in Fig. 3a. The results show that the 'one filled, one empty' configuration is strongly favored over a wide range of magnetic field centered to $B/B^{\Phi}=1$ for the short-range α, β pairs (Fig. 3d and 3e, respectively). In contrast, the favorable and frustrated pairwise populations for the long range γ pairs (see Fig. 3b for definition) nearly overlap with those for the non-interactive pairs over the entire magnetic field range, indicating uncorrelated long range γ pairs. We found that the repulsive potential centers (red circles in Fig. 2 and 3b) play an important role in breaking down the γ pairwise correlation, as indicated by the appearance of highly favorable γ pairs and the emergence of long range ordered FQs when the repulsive potential centers are removed (see Extended Data Fig. 6). Thus, the frustrated short range correlations and the breakdown of long range order, respectively governed by attractive and repulsive potential centers, lead to a massive $2^n$-fold degeneracy (with n being the number of square plaquettes) in the entire 2D square lattice.

7. Since the FQ density at $B/B^{\Phi}=1$ corresponds to one FQ per plaquette, equal numbers of the two different frustrated plaquettes ('two filled' and 'two empty' α pairs) must emerge simultaneously. These frustrated plaquettes can be assigned a positive ('two filled' α pair) or negative ('two empty' α pair) net FQ with respect to the background of one FQ per plaquette, resembling the simultaneous emergence of positive and negative magnetic monopole pairs connected by Dirac strings in a magnetic spin-ice system [28]. Therefore, 'Dirac strings' connecting pairs of 'two filled' and 'two empty' frustrated plaquettes can also be designed in our FQ system as demonstrated in Extended Data Fig. 7. Although the emergence of frustrated plaquettes (frustrated α pairs) reduces the degenerate manifold, the location of these defects provides a new degree of freedom, thus inducing massively degenerate states for excited defects.

8. Figure 2a indicates two ways of switching FQs between geometrically frustrated and crystallized states: 1) tuning FQ density by magnetic field and 2) switching potential landscape by reconfiguring MC orders. The first method is unique to the FQ system, enabling easily controlled defect formation and tunable geometric frustration by the external magnetic field (Extended Data Fig. 8). Unlike the recent realization of a thermally controlled potential landscape for pinning the FQs [18], which is only accessible in superconducting systems, our second method of magnetically switching the potential landscapes with reconfigurable MC orders could be applied to other magnetically interacting particle systems, such as magnetic skyrmions [6] and magnetic colloids [10,11].

9. The ability to tune the symmetry of the FQ lattice in-situ (Fig. 2) can be extended to design superconducting devices such as a FQ rectifier. Such a FQ diode can be realized based on the ratchet motion of FQs induced by an asymmetric potential [29-31]. We applied an AC current to the superconducting layer of the hetero-structure, and measured a DC voltage response associated with the directional motion of FQs (Fig. 4a,). The rectification effect is observed for the Type-II (Fig. 4c) and Type-III (Fig. 4e) MC configuration and can be turned off with the Type-I (Fig. 4b) configuration. The Supplementary Videos S1, S2 and S3 clearly show the asymmetric dynamics of FQs motion under Type-II and Type-III MC orders, respectively. We can also reverse the rectification polarity by reversing the MC configurations, as shown in Fig. 4d and 4f for Type-II and Type-III configurations, respectively. This reprogrammable spin-ice FQ diode could provide the basis of logic gates for low power computing.

10. In addition to designing geometric frustration in various magnetic systems, reconfigurable electronic devices may also be envisioned by combining an artificial-spin-ice with functional materials, such as 2D electron gases [7], graphene [8], topological insulators [9], and Dirac and Weyl semimetals. Furthermore, the precise control of the MC configuration with a magnetic tip [5] enables local 'write and erase' of geometric frustrated and crystallized patterns, leading to even more exotic properties and an enhanced control of functionalities at the nanoscale. The distinguishable features in the transport behavior shown in Fig. 1f-1h and Extended Data Fig. 2 demonstrate that electrical transport measurements on a subfilm is a convenient technique to detect the spin/charge configurations of magnetic systems and complement the magnetic imaging techniques. Finally, by replacing the present Type-I MCs with nano-bar magnets or magnetic nanodots (Extended Data Fig. 9), one can create novel geometric frustration that can be investigated by direct imaging and thermal annealing techniques [23].

**References:**

[1] A. P. Ramirez, "Geometric frustration: Magic moments", *Nature,* **421**, 483 (2003)

[2] L. J. Heyderman and R. L. Stamps, "Artificial ferroic systems: novel functionality from structure, interactions and dynamics", *J. Phys. Conden. Matter,* **25**, 363201 (2013)

[3] Y. Perrin, B. Canals and N. Rougemaille, "Extensive degeneracy, Coulomb phase and magnetic monopoles in artificial square ice", *Nature,* **540**, 410 (2016)

[4] E. H. Lieb, "Residual entropy of square ice", *Phys. Rev.,* **162**, 162 (1967)

[5] Y.-L. Wang, Z.-L. Xiao, A. Snezhko, J. Xu, L. E. Ocola, R. Divan, J. E. Pearson, G. W. Crabtree and W.-K. Kwok, "Rewritable artificial magnetic charge ice", *Science,* **352**, 962 (2016)

[6] F. Ma, C. Reichhardt, W. Gan, C. J. O. Reichhardt and W. S. Lew, "Emergent geometric frustration of artificial magnetic skyrmion crystals", *Phys. Rev. B. ,* **94**, 144405 (2016)

[7] A. Singha, M. Gibertini, B. Karmakar, S. Yuan, M. Polini, G. Vignale, M. I. Katsnelson, A. Pinczuk, L. N. Pfeiffer, K. W. West and V. Pellegrini, "Two-Dimensional Mott-Hubbard Electrons in an Artificial Honeycomb Lattice", *Science,* **332**, 1176 (2011)

[8] M. Taillefumier, V. K. Dugaev, B. Canals, C. Lacroix and P. Bruno, "Graphene in a periodically alternating magnetic field: An unusual quantization of the anomalous Hall effect", *Phys. Rev. B,* **84**, 085427 (2011)

[9] I. Gilbert, C. Nisoli and P. Schiffer, "Frustration by design", *Phys. Today,* **69**, 54 (2016)

[10] J. Loehr, A. Ortiz-Ambriz and P. Tierno, "Defect Dynamics in Artificial Colloidal Ice: Real-Time Observation, Manipulation, and Logic Gate", *Phys. Rev. Lett.,* **117**, 168001 (2016)

[11] A. Ortiz-Ambriz and P. Tierno, "Engineering of frustration in colloidal artificial ices realized on microfeatured grooved lattices", *Nat. Commun.,* **7**, 10575 (2016)

[12] Y. Han, Y. Shokef, A. M. Alsayed, P. Yunker, T. C. Lubensky and A. G. Yodh, "Geometric frustration in buckled colloidal monolayers", *Nature,* **456**, 898, (2008)

[13] A. Lib\'al, C. Reichhardt and C. J. Olson, "Realizing Colloidal Artificial Ice on Arrays of Optical Traps", *Phys. Rev. Lett.,* **97**, 228302 (2006)

[14] P. Gammel, "Why vortices matter", *Nature,* **411**, 434 (2001)

[15] S. Savel'EV and F. Nori, "Experimentally realizable devices for controlling the motion of magnetic flux quanta in anisotropic superconductors", *Nat. Mater.,* **1**, 179 (2002)

[16] A. Lib\'al, C. J. Olson and C. Reichhardt, "Creating Artificial Ice States Using Vortices in Nanostructured Superconductors", *Phys. Rev. Lett.,* **102**, 237004 (2009)

[17] M. L. Latimer, G. R. Berdiyorov, Z. L. Xiao, F. M. Peeters and W. K. Kwok, "Realization of Artificial Ice Systems for Magnetic Vortices in a Superconducting MoGe Thin Film with Patterned Nanostructures", *Phys. Rev. Lett.,* **111**, 067001 (2013)

[18] TrastoyJ., MalnouM., UlysseC., BernardR., BergealN., FainiG., LesueurJ., BriaticoJ. and J. E. Villegas, "Freezing and thawing of artificial ice by thermal switching of geometric frustration in magnetic flux lattices", *Nat. Nano.,* **9**, 710 (2014)

[19] C. Nisoli, R. Moessner and P. Schiffer, "Colloquium: Artificial spin ice: Designing and imaging magnetic frustration", *Rev. Mod. Phys.,* **85**, 1473 (2013)

[20] R. F. Wang, C. Nisoli, R. S. Freitas, J. Li, W. McConville, B. J. Cooley, M. S. Lund, N. Samarth, C. Leighton, V. H. Crespi and P. Schiffer, "Artificial 'spin ice' in a geometrically frustrated lattice of nanscale ferromagnetic islands", *Nature,* **439**, 303 (2006)

[21] C. Nisoli, J. Li, X. Ke, D. Garand, P. Schiffer and V. H. Crespi, "Effective Temperature in an Interacting Vertex System: Theory and Experiment on Artificial Spin Ice", *Phys. Rev. Lett.,* **105**, 047205 (2010)

[22] A. Farhan, P. M. Derlet, A. Kleibert, A. Balan, R. V. Chopdekar, M. Wyss, J. Perron, A. Scholl, F. Nolting and L. J. Heyderman, "Direct Observation of Thermal Relaxation in Artificial Spin Ice", *Phys. Rev. Lett.,* **111**, 057204 (2013)

[23] S. Zhang, I. Gilbert, C. Nisoli, G.-W. Chern, M. J. Erickson, L. O'Brien, C. Leighton, P. E. Lammert, V. H. Crespi and P. Schiffer, "Crystallites of magnetic charges in artificial spin ice", *Nature,* **500**, 553 (2013)

[24] J. P. Morgan, A. Stein, S. Langridge and C. H. Marrows, "Thermal ground-state ordering and elementary excitations in artificial magnetic square ice", *Nat. Phys.,* **7**, 75 (2011)

[25] V. Kapaklis, U. B. Arnalds, A. Farhan, R. V. Chopdekar, A. Balan, A. Scholl, L. J. Heyderman and B. Hjorvarsson, "Thermal fluctuations in artificial spin ice", *Nat. Nano.,* **514**, 9 (2014)

[26] G. Moller and R.Moessner, "Artificial Square Ice and Related Dipolar Nanoarrays", *Phys. Rev. Lett.,* **96**, 237202 (2006)

[27] M. V. Milosevic and F. M. Peeters, "Vortex pinning in a superconducting film due to in-plane magnetized ferromagnets of different shpes: The London approximation", *Phys. Rev. B,* **69**, 104522 (2004)

[28] E. Mengotti, L. J. Heyderman, A. F. Rodriguez, F. Nolting, R. V. Hugli and H.-B. Braun, "Real-space observation of emergent magnetic monopoles and associated Dirac strings in artificial kagome

spin ice", *Nat. Phys.,* **7**, 68 (2011)

[29] J. E. Villegas, S. Savelev, F. Nori, E. M. Gonzalez, J. V. Anguita, R. Garcia and J. L. Vicent, "A superconducting reversible rectifier that controls the motion of magnetic flux quanta", *Science,* **302**, 1188 (2003)

[30] C. C. d. S. Silva, J. V. d. Vondel, M. Morelle and V. V. Moshchalkov, "Controlled multiple reversals of a ratchet effect", *Nature,* **440**, 651 (2006)

[31] C. S. Lee, B. Janko, I. Derenyi and A. L. Barabasi, "Reducing vortex density in superconductors using the 'ratchet effect' ", *Nature,* **400**, 337 (1999)

**Acknowledgements:** This work was supported by the U.S. Department of Energy, Office of Science, Basic Energy Sciences, Materials Sciences and Engineering Division. Use of the Center for Nanoscale Materials, an Office of Science user facility, was supported by the U. S. Department of Energy, Office of Science, Office of Basic Energy Sciences, under Contract No. DE-AC02-06CH11357. Z.L.X. and J.X. acknowledge NSF Grant No. DMR-1407175.

**Figure 1**

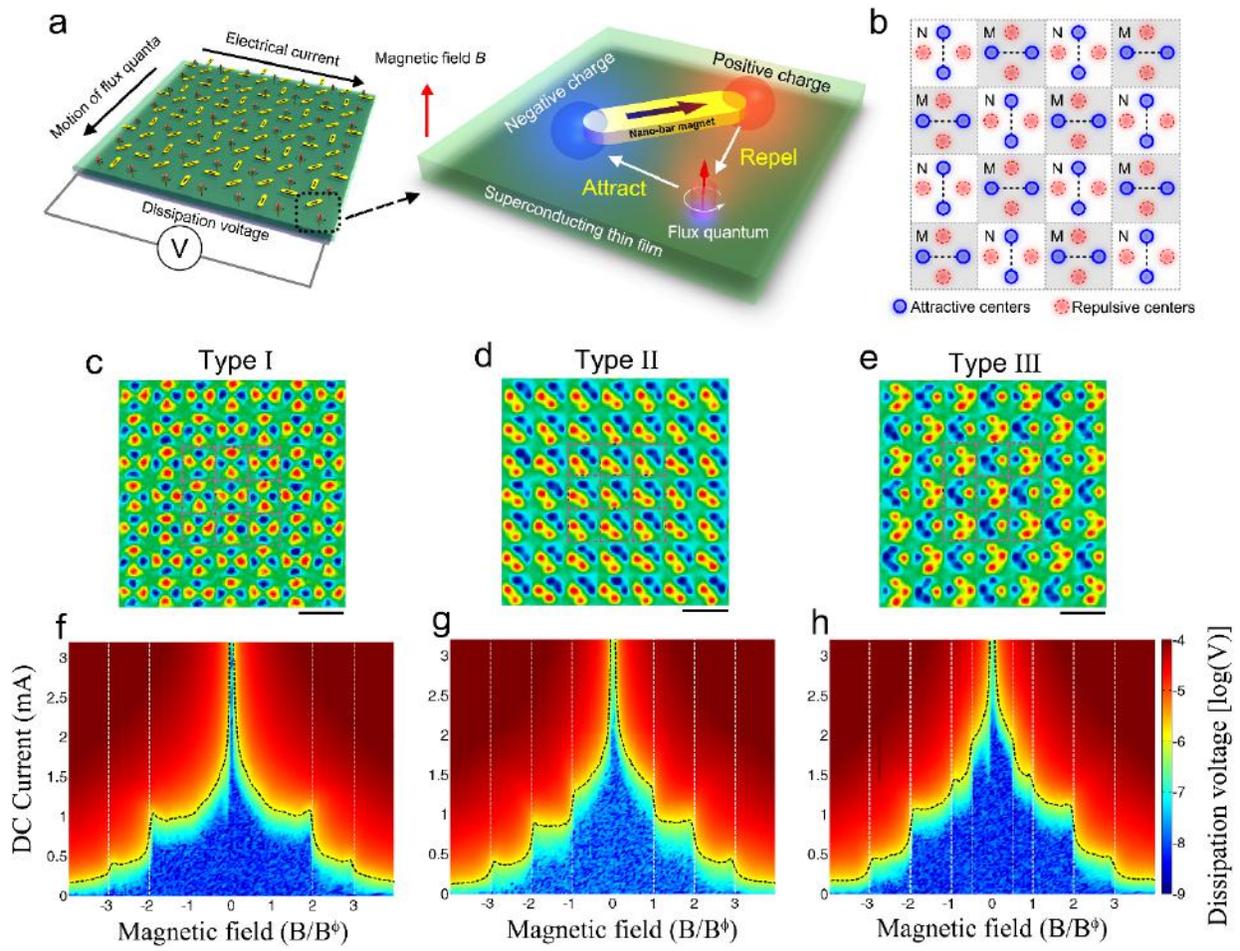

**Figure 1 | Reconfigurable potential landscapes and flux-quantum motion controlled by artificial-spin-ice. a**, Hetero-structure consisting of a reconfigurable artificial-spin-ice (interacting nanoscale bar magnets) on top of a superconducting thin film. Dashed arrow points to a magnified image delineating the interaction between a flux quantum (pointing 'up') in the superconducting layer and the magnetic dipole charges. The attractive and repulsive interactions for a flux quantum will be reversed for a flux-quantum pointing 'down'. **b**, Two sublattices of the Type-I magnetic charges. The red and blue circles represent positive and negative magnetic charges, respectively. **c-e**, Magnetic force microscopy images of three typical magnetic charge orders obtained from the fabricated artificial-spin-ice pattern (Extended Data Fig. 1a) of the heterostructure. The dashed lines show the 3×3 square lattice of the charges. Scale bar, 1 µm. **f-h**, Color maps of the magnetic field and current dependent voltages associated with magnetic charge orders shown in (**c-e**), respectively. The black-dashed curves indicate the superconducting critical currents defined by a voltage criterion of 1µV. The magnetic field unit $B^\Phi$=40 Oe corresponds to a flux density of one flux quantum per square plaquette.

Figure 2

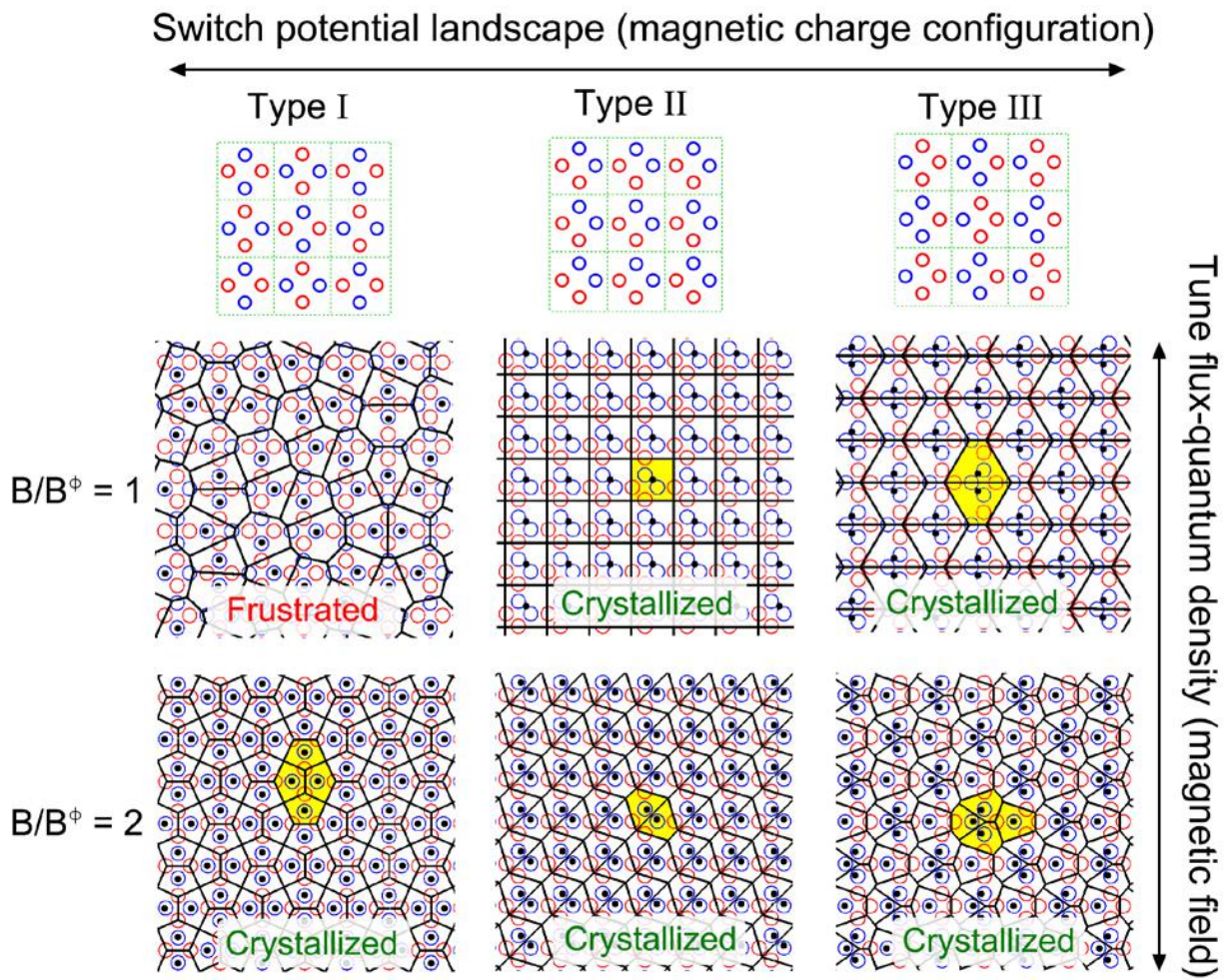

**Figure 2 | Switchable geometric frustration and crystallization.** Simulated distributions of flux quanta under Type-I, II and III magnetic charge order and various flux quantum density. Solid black lines constitute Voronoi diagrams elucidating the flux quanta ordering. One repeating unit structure in each of the crystallized flux-quanta lattice is highlighted in yellow. The disordered flux-quanta under Type-I magnetic charge configuration at $B/B^{\Phi}=1$ originate from geometric frustration with extensive degeneracy.

Figure 3

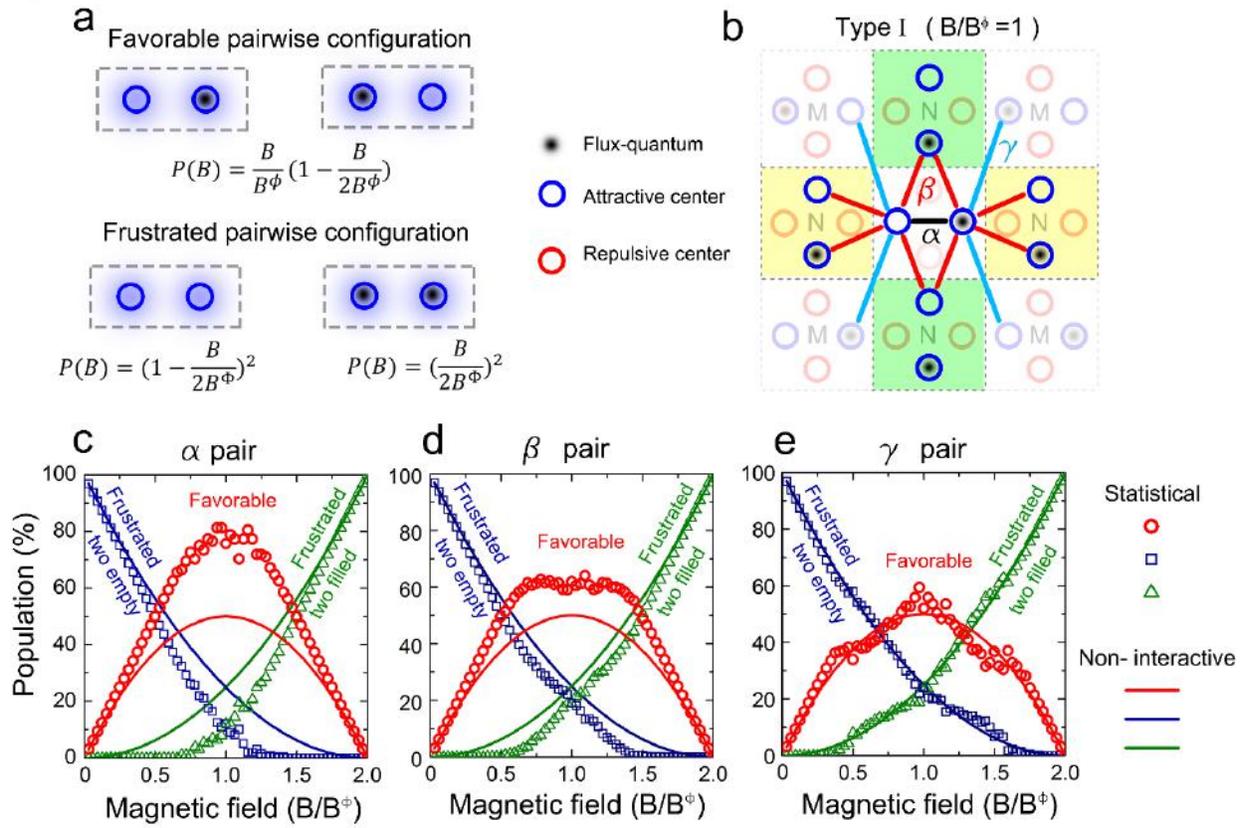

**Figure 3 | Geometrically frustrated local interactions and breakdown of long range order.**
**a**, Flux quantum filling configurations for a pair of attractive potential centers. The magnetic field dependent population rate $P(B)$ of each configuration for non-interactive flux quanta is listed below each configuration, where $B^{\Phi}=40$ Oe corresponds to one flux quantum per plaquette (each plaquette contains two attractive potential centers). **b**, Illustration of α, β and γ pairs of attractive centers under Type-I charge order. Switching the flux-quantum between the two attractive centers in a M (or N) plaquette does not change the interaction energy with the two green N (or M) plaquettes, and cannot simultaneously minimize the energy with the two yellow N (or M) plaquettes. This leads to a two-fold degeneracy in each plaquette. **c-e**, Magnetic field dependent populations (open symbols) of pairwise flux quantum filling configurations. The solid lines represent populations for noninteracting flux quanta calculated from the formula listed in (a).

Figure 4

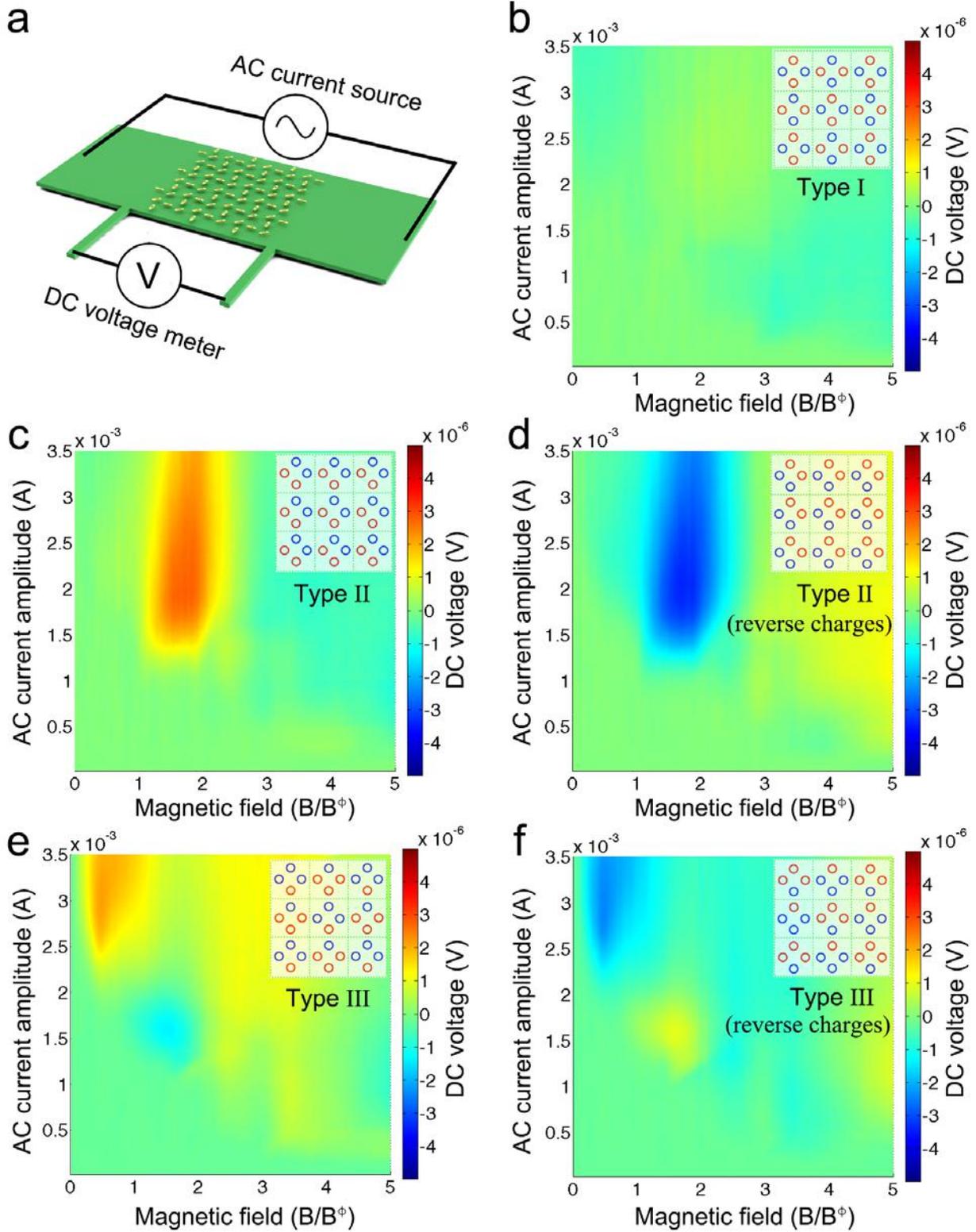

Figure 4: **Reprogrammable flux quantum rectifier. a**, Schematic of the experimental setup with AC current drive and DC voltage detection. **b-f**, Color maps of DC voltage response associated with rectification as a function of AC current amplitude and magnetic field under Type I, II and III magnetic charge order shown in the insets. The magnetic charges in (**d**) and (**f**) are reversed (positive to negative and vice versa) as compared to those in (**c**) and (**e**), respectively.

**Method**

**Sample fabrication.** A superconducting $Mo_{0.79}Ge_{0.21}$ (MoGe) film of 50 nm thick was sputtered from a MoGe alloy target onto silicon substrates with an oxide layer. The samples were first patterned into a 50 μm-wide micro-bridge with voltage leads separated by 50 μm using standard photolithography. Subsequently, an artificial-spin-ice of permalloy ($Ni_{0.80}Fe_{0.20}$) bar magnets (each with dimension: 300 nm × 80 nm × 25 nm) was patterned on top of the MoGe film using electron-beam lithography. The detailed fabrication parameters and procedures for patterning the tri-axial ASI can be found in reference [5]. The MoGe film has a superconducting transition temperature of 6.9 K, as shown in Extended Data Fig. 10.

**Experiments**. Magnetic force microscopy (MFM) imaging was conducted in a custom designed MFM system using a commercial MFM probe (NANOSENSORS™ PPP-MFMR). The detailed imaging parameters can be found in reference [5]. The transport experiments were carried out using a standard four-probe method at a temperature of 5.8 K. The sample was placed in a 3-axis superconducting vector magnet, which allows us to apply magnetic field in any desired orientation. The MC ordering was manipulated by controlling both the orientations and amplitude of the in-plane magnetic field. The detailed protocols can be found in reference [5]. The transport experiments were carried out with only the out-of-plane magnetic field. In-plane magnetic field was only used to reconfigure the MC state and ramped to zero thereafter.

**Molecular dynamic simulation.** We consider a two-dimensional system with periodic boundary condition in x and y direction. The sample size is $L \times L$ with $L=16\lambda$. $\lambda$ is the London penetration depth, which is chosen as 720nm for MoGe [17]. The MCs at the bar magnet ends are kept as attractive or repulsive centers for FQ. The interaction energy of a FQ with a single pinning site is modelled as a Gaussian function, with attractive (repulsive) pinning sites introducing negative (positive) interaction energy to the system.

The dynamics of an individual FQ $i$ is determined by the following overdamped equation of motion:

$$\eta \frac{d\vec{R}_i}{dt} = \vec{F}_i^{vv} + \vec{F}_i^{pa} + \vec{F}_i^{pr} + \vec{F}^d + \vec{F}_i^T$$

Here $\eta$ is the damping constant which is set equal to 1. The repulsive inter-FQs interaction force is given by $\vec{F}_i^{vv} = \sum_{j \neq i} F_0 K_1(R_{ij}/\lambda)\hat{R}_{ij}$, where $\vec{R}_i$ is the location of FQ $i$, $K_1$ is the modified Bessel function, $R_{ij} = |\vec{R}_i - \vec{R}_j|$, $\hat{R}_{ij} = (\vec{R}_i - \vec{R}_j)/R_{ij}$, $F_0 = \phi_0^2/(2\pi\mu_0\lambda^3)$, $\phi_0$ is the flux quantum, and $\mu_0$ is the permeability [32]. The pinning force is derived from the gradient of pinning potential energy, with the form

$$\vec{F}_i^{pa} = -\sum_{k=1}^{N_{pa}} F_p R_{ik}^{(pa)} \exp(-R_{ik}^{(pa)^2}/r_p^2) \hat{R}_{ik}^{(pa)},$$

where $r_p = 240 nm$ is the standard deviation of Gaussian function, $F_p$ is the pinning coefficient, $\vec{R}_k^{(pa)}$ is the center of attractive pinning site $k$, $R_{ik}^{(pa)} = |\vec{R}_i - \vec{R}_k^{(pa)}|$, and $\hat{R}_{ik}^{(pa)} = (\vec{R}_i - \vec{R}_k^{(pa)})/R_{ik}^{(pa)}$. $\vec{F}_i^{pr}$ has the same form as $\vec{F}_i^{pa}$, but the opposite sign. $\vec{F}^d$ is the Lorentz force when there is a constant current applied, following $\vec{F}^d = (\vec{J} \times \hat{z})\phi_0 d$ [28], where $\vec{J}$ is the current density, $\hat{z}$ is the direction of FQ vorticity, and $d$ is the thickness of the superconducting film. Thermal forces are modeled as Langevin kicks $\vec{F}_i^T$ satisfying $<\vec{F}_i^T(t)> = 0$ and $<\vec{F}_i^T(t) \cdot \vec{F}_j^T(t')> = 2\eta k_B T \delta_{ij} \delta(t-t')$, where $k_B$ is the Boltzmann constant.

We start from a high temperature where FQs are in a molten state and then slowly reduce the temperature to zero with 130 intermediate temperatures. At each temperature step, we let the system evolve $2 \times 10^5$ steps, each being $dt = 0.01$ in simulation unit. After annealing, we obtain the static FQ configurations as the ground states. To obtain the critical current, we apply and increase the driving force slowly. For each driving force, we measure the time-averaged total FQ velocity in the force direction, $V_x = \frac{1}{N_t}\sum_t^{N_t}\sum_i^{N_v} \hat{x} \cdot \vec{v}_i$. $V_x$ corresponds to the dissipation voltage. When $V_x$ exceeds the threshold which is set at 40 in simulation, most FQs start to move and the non-dissipative superconducting phase breaks down. The critical driving force is defined as the depinning force, which represents the critical current. To speed up our simulation, we used binary search here to find the critical current. Although the true ground states may only be attainable for prohibitively long waiting times, the agreement between our simulation results and experiments indicates validation of our model [33].

[32] C. Reichhardt, D. Ray, and C. J. Olson Reichhardt, "Reversible ratchet effects for vortices in conformal pinning arrays", *Phys. Rev. B.,* **91**, 184502 (2015)

[33] C. Reichhardt, N. Gronbech-Jensen, "Critical currents and vortex states at fractional matching fields in superconductors with periodic pinning", *Phys. Rev. B.,* **63**, 054510 (2001)

**Extended Data Figures**

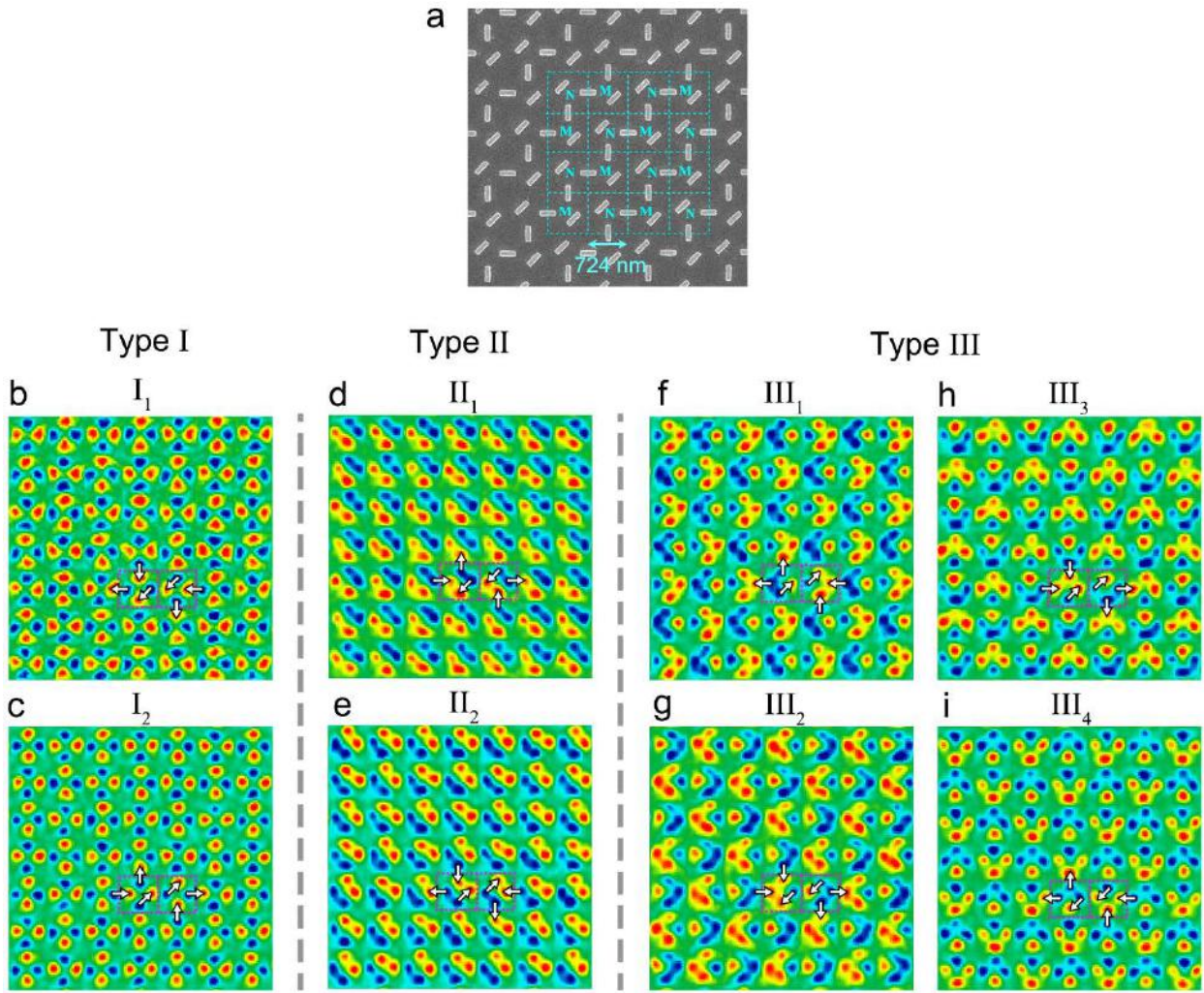

**Extended Data Figure 1 | All eight possible configurations of long range magnetic charge ordering. a,** Scanning electron microscopy image of the tri-axial artificial-spin-ice patterned on top of a MoGe film. Two sublattices are labeled with M and N, respectively. The square lattice constant is about 724 nm. The dimension of the nano-bar magnets is 300 nm×80 nm×25 nm. **b-i,** Magnetic force microscopy images of eight configurations of ordered magnetic charges obtained from the tri-axial artificial-spin-ice structure (a) in our hetero-structure. The red and blue spots denote positive and negative magnetic charges, respectively. The purple boxes outline the M and N plaquettes defined in Fig. 1b. The arrows indicate the spin configuration of the nanoscale bar magnets.

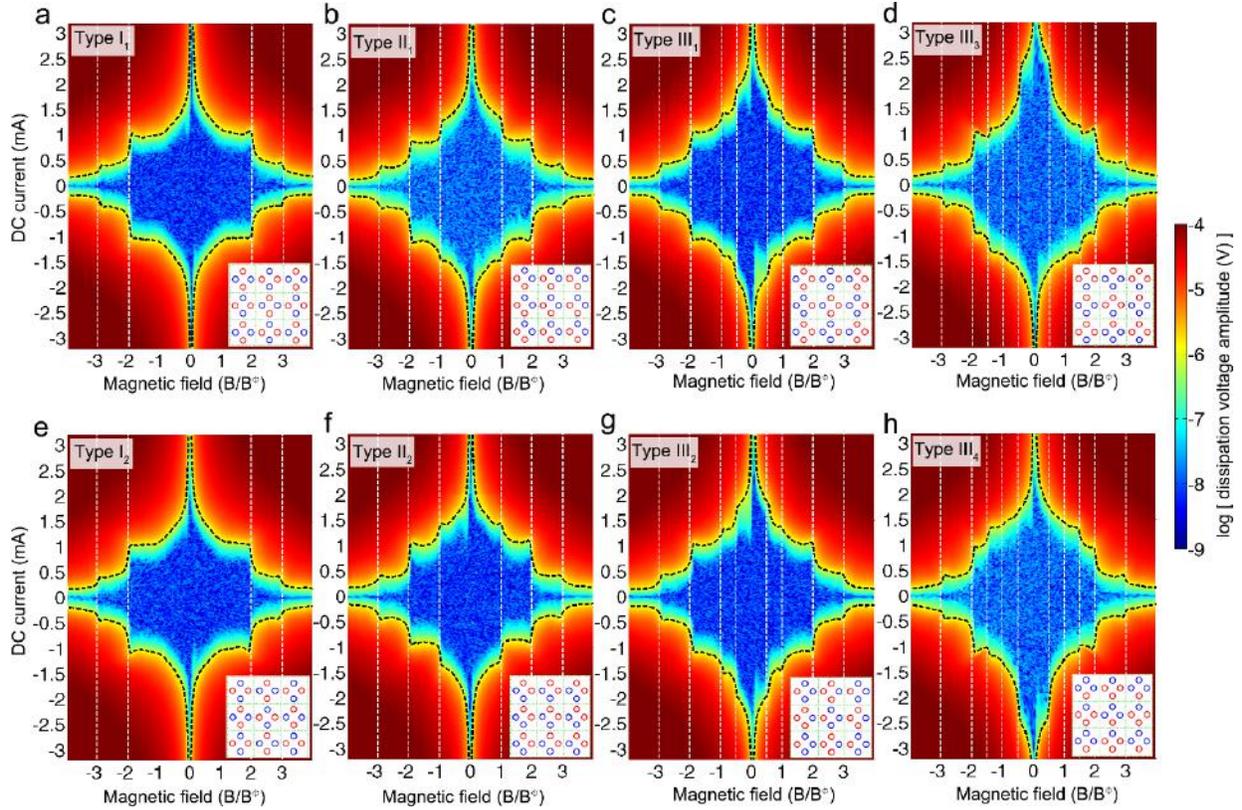

**Extended Data Figure 2 | Magnetic field matching effect for all eight possible configurations of magnetic charges**. Color maps of the magnetic field and current dependent voltage. The corresponding magnetic charge configurations are presented in the insets. The current is in the horizontal direction. The magnetic field is perpendicular to the sample surface. The unit $B^\Phi$=40 Oe corresponds to one flux quantum per square plaquette. The black dashed lines indicate the magnetic field dependent critical current curves defined by a voltage criterion of 1 µV. The voltage maps are symmetric with magnetic field polarity for all the magnetic charge configurations. They are symmetric with current polarity for Type $I_1$ (a), $I_2$ (e), $III_1$ (c) and $III_2$ (g) charge configurations and asymmetric with current polarity for Type $II_1$ (b), $II_2$ (f), $III_3$ (d) and $III_4$ (h) charge configurations. The magnetic charge polarities of Type $I_1$, $II_1$, $III_1$ and $III_3$ (a-d) are reversed (positive charges to negative charges and vice versa) as compared to those of Type $I_2$, $II_2$, $III_2$ and $III_4$ (e-h), respectively. This leads to the results at positive current in (a-d) equal to those at negative current (e-h) respectively and vice versa (The flux-quanta move oppositely for positive and negative currents).

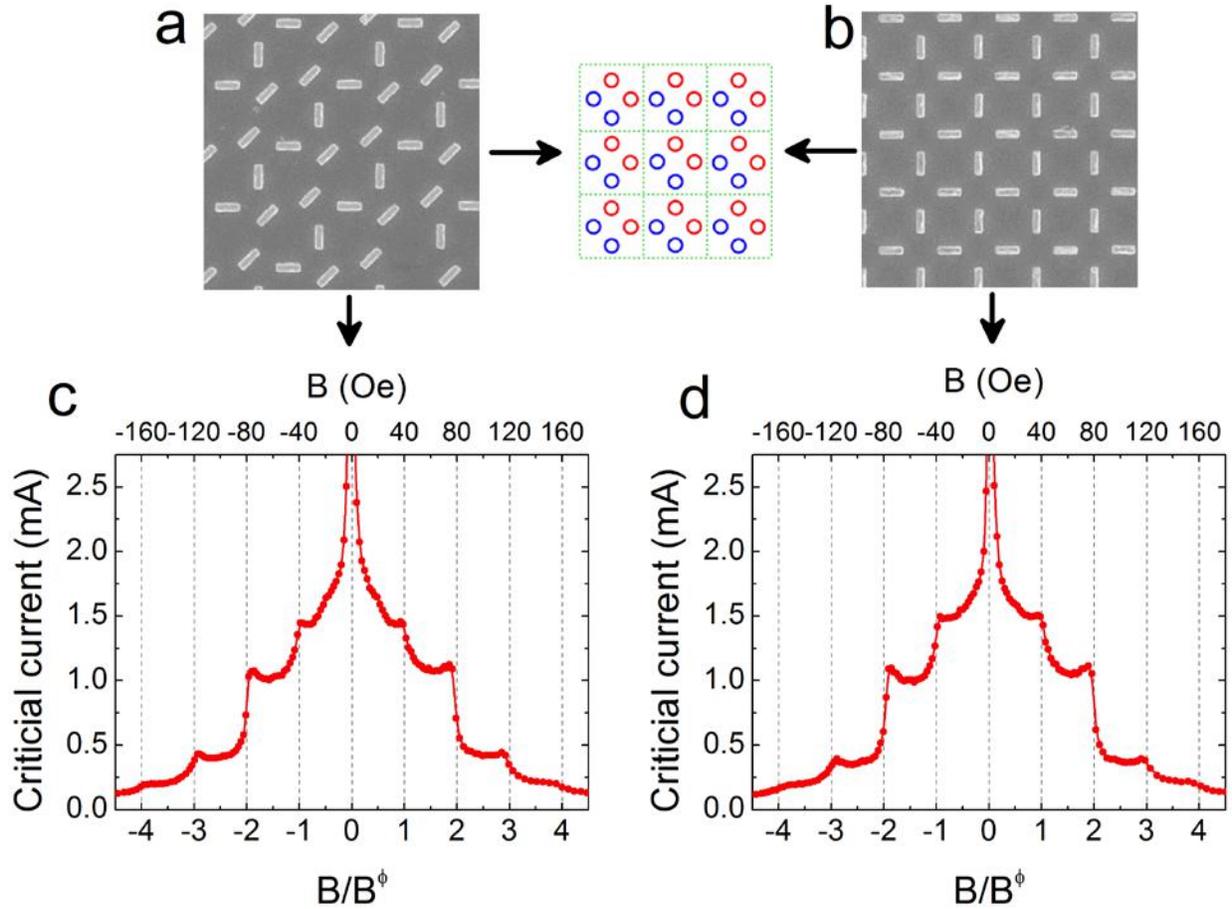

**Extended Data Figure 3 | Magnetic charge driven flux-quanta matching effect. a,b,** Scanning electron microscopy images of our tri-axial artificial-spin-ice structure (**a**) which enables in-plane field control of the three sub-latttice oriented nanomagnets compared with a typical square artificial-spin-ice structure (**b**) in which Type-II magnetic charge order can be obtained by diagonally polarized in-plane magnetic field. **c,d**, Magnetic field dependent critical current curves measured from tri-axial artificial-spin ice (**c**) and square artificial-spin-ice (**d**) on a superconducting film under identical Type-II magnetic charge configurations shown in the middle. Both artificial-spin-ice structures show nearly identical matching effects.

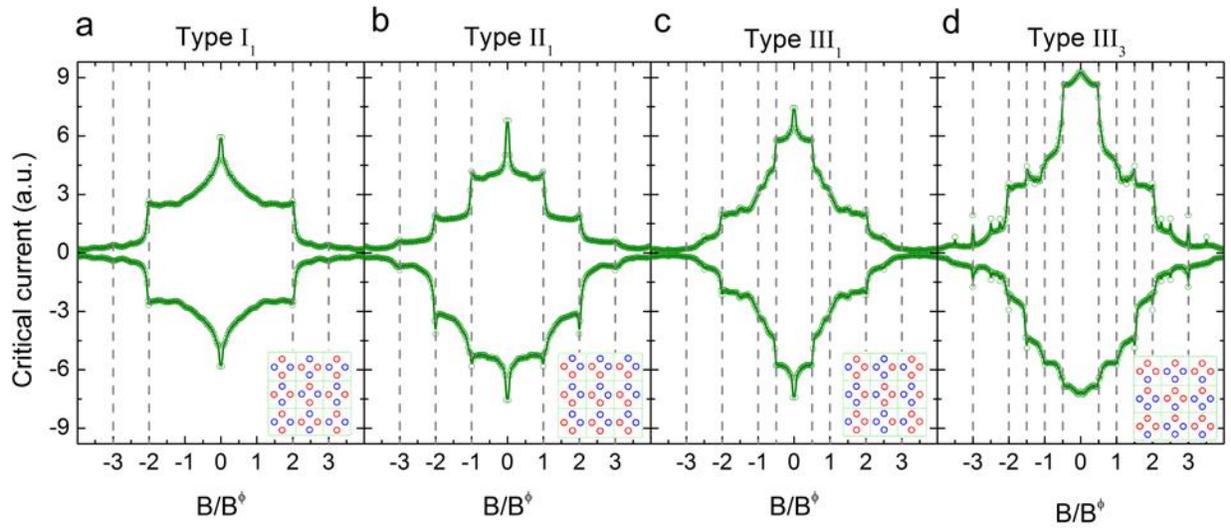

**Extended Data Figure 4 | Simulated magnetic field dependent critical currents for Type I, II and III magnetic charge order configuration. a-d**, corresponds to experimental results in Extended Data Figure 2a-2d, respectively. The insets show the associated magnetic charge configurations.

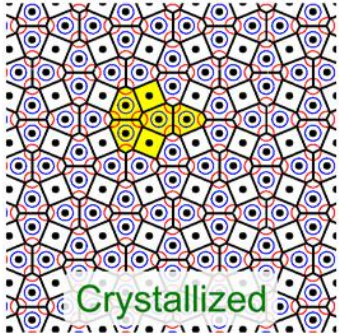
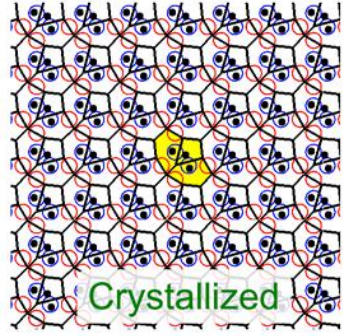
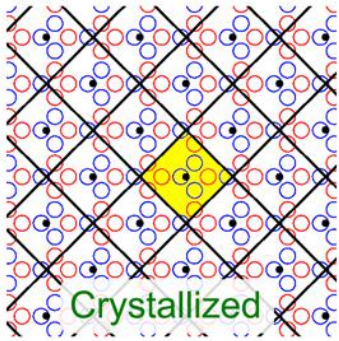
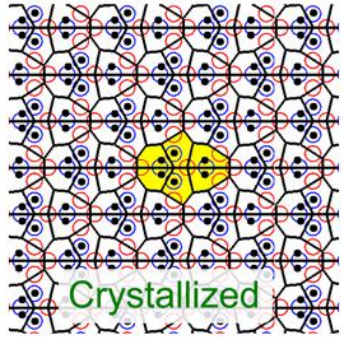

**Extended Data Figure 5 | Several more simulated distributions of flux-quanta.** The magnetic charge configurations and magnetic field are listed above each image.

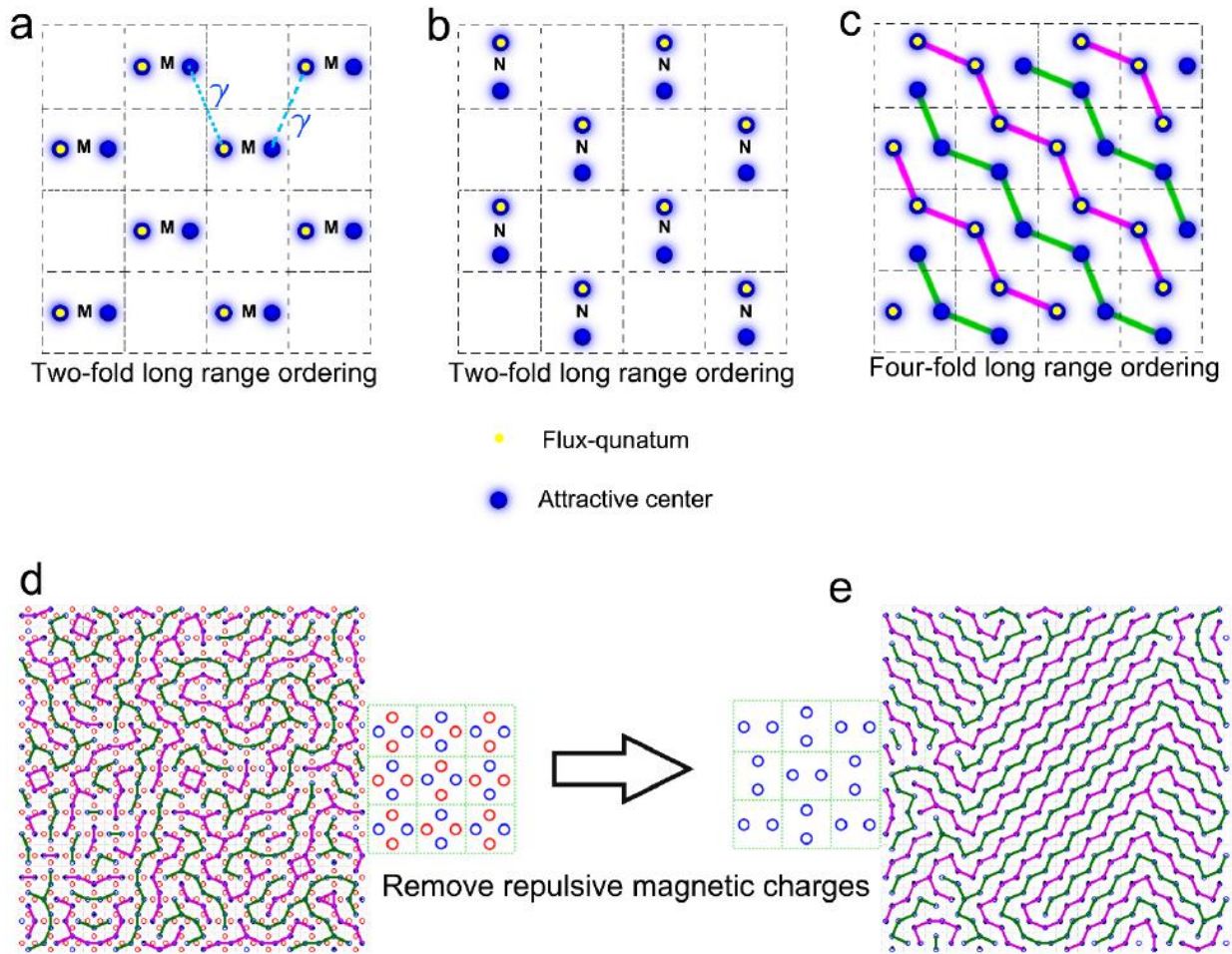

**Extended Data Figure 6 | Emergence of long range ordered flux-quanta when the repulsive potential centers are removed. a,b**, sublattices containing only M plaquettes (a) or N plaquettes (b). The favorable 'one filled, one empty' configuration can be satisfied for all γ pairs when the flux-quanta all fill either the left or right attractive centers in each M plaquette (top or bottom attractive centers in each N plaquette), leading to a two-fold long range ordering of flux-quanta in each sublattice. **c**, four-fold long range ordering of flux-quanta in a full lattice (both M and N plaquettes) satisfies all γ pairs in the favorable 'one filled, one empty' configuration. The Labyrinth patterns obtained by connecting frustrated short-range α and β pairs of 'two filled' (purple) and 'two empty' (green) show stripe-like long range ordering of flux-quanta. **d,e**, The disordered flux-quanta (random Labyrinth patterns) induced by Type-I charge order with both attractive and repulsive centers transform into long range ordered flux-quanta (large domains of stripe-like Labyrinth patterns) when the repulsive centers are removed.

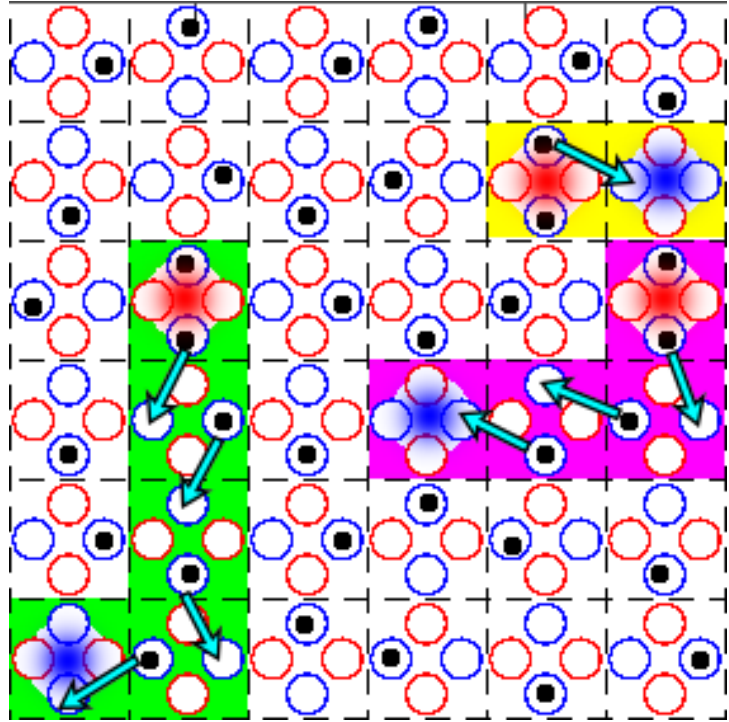

**Extended Data Figure 7 | Flux-quanta 'monopoles' and 'Dirac strings'.** The frustrated α pair in each square plaquette has a net flux-quantum 'charge' of +1 and -1 for 'two filled' and 'two empty' α pairs, respectively, as highlighted by the red and blue central spots. Three Dirac strings connecting a pair of positive and negative flux quanta 'charges' are shown in green, yellow and purple colors, respectively. The arrows show possible flux-quanta motion leading to annihilation of the positive and negative flux quanta 'charges'.

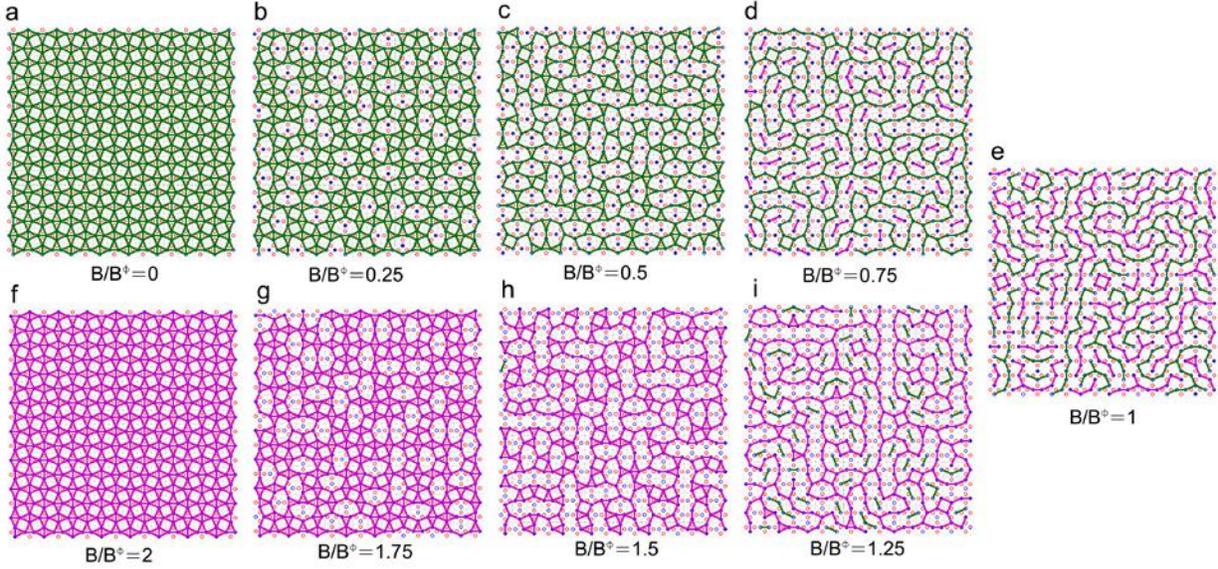

**Extended Data Figure 8 | Evolution of magnetic field dependent geometric frustration and crystallization under Type I charge order.** Labyrinth patterns obtained by connecting short-range frustrated α and β pairs (see Fig. 3c for pair definition) 'two filled' (purple) and 'two empty' (green) show distributions of the flux-quanta and the competing energies (green: FQ/MC interaction energy; purple: inter-FQs interaction energy) at various magnetic fields (or flux-quanta density) listed below each figure. The number of filled (empty) attractive centers in the top panels (a-d) equal to that of the empty (filled) attractive centers in the bottom panels (f-i), respectively. The $B/B^{\Phi}=1$ pattern shows highly frustrated configuration while the $B/B^{\Phi}=0$ and 2 show perfectly ordered patterns.

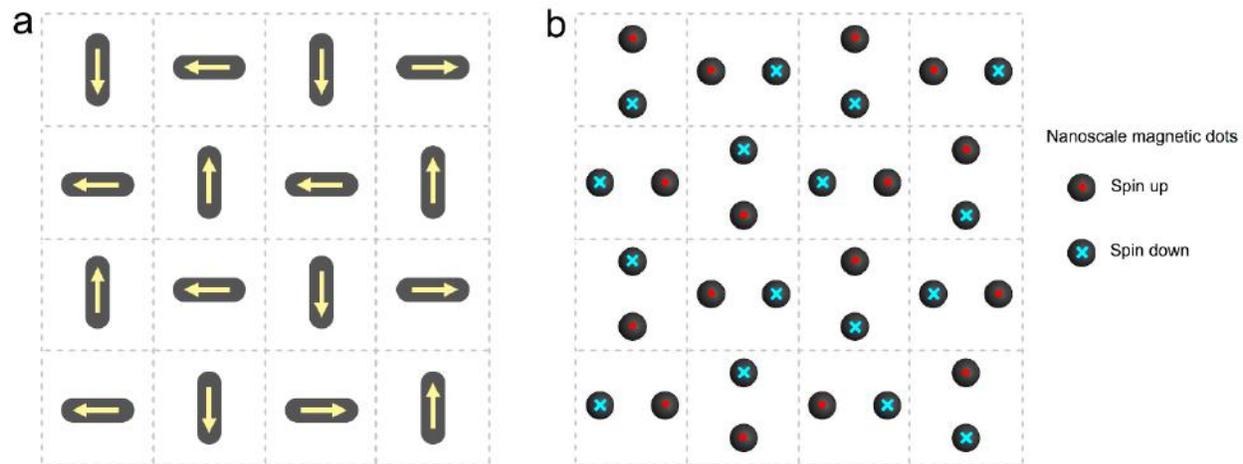

**Extended Data Figure 9 | Artificial-spin-ice analog of the new geometric frustration in our flux-quantum system. a,** an array of nanoscale bar magnets. **b**, an array of nanoscale magnetic dots with out-of-plane anisotropy.

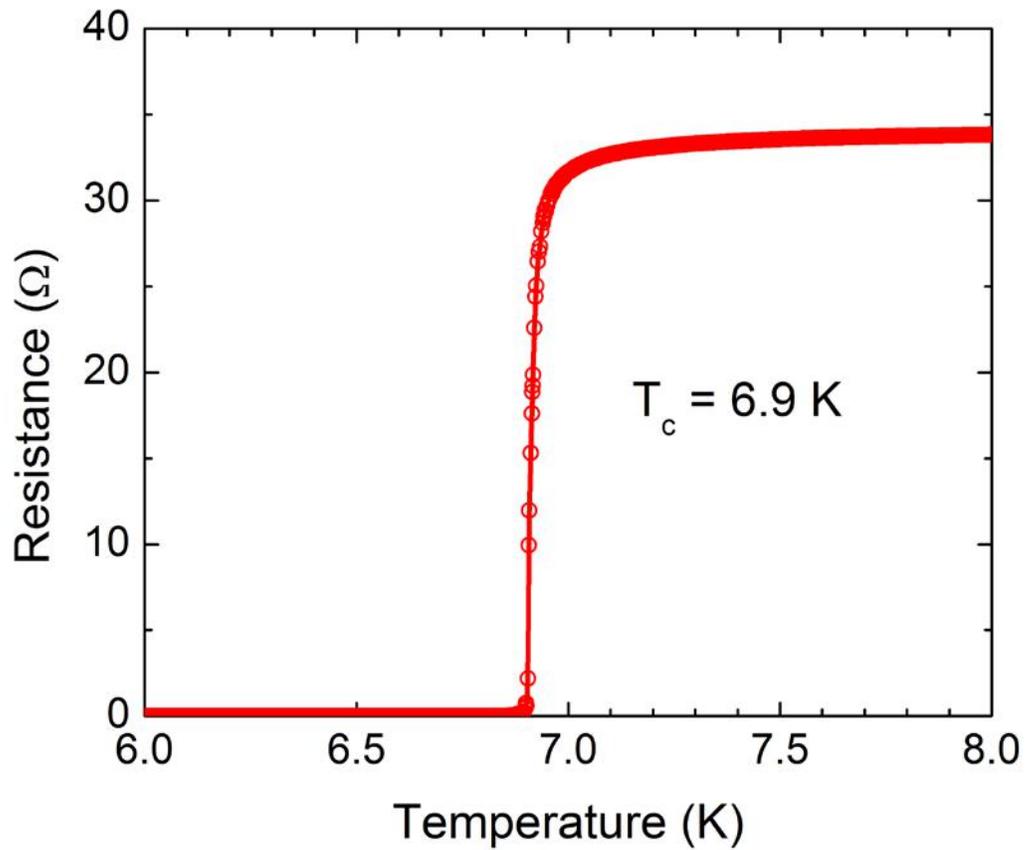

**Extended Data Figure 10 | Temperature dependent resistance.** The resistance versus temperature curve displays a sharp superconducting transition with a critical temperature of $T_c \approx$ 6.9 K. The resistance was measured with an applied current of 100 µA under zero external magnetic field.

**Video 1 | DC Transport Property under Type-II MC Configuration at $B/B^\Phi = 1.5$**

Video showing asymmetric DC transport behavior of FQs under Type-II MC configuration at $B/B^\Phi = 1.5$. With the same current amplitude, FQs move more easily in one direction than the other. This leads to ratchet motion of FQs as well as a DC dissipation voltage response under AC driving current. FQs are represented by black dots. Color map shows relative amplitude of pinning potential, with positive for repulsive potentials and negative for attractive potentials.

**Video 2 | DC Transport Property under Type-III MC Configuration at $B/B^\Phi = 0.5$**

Video showing asymmetric DC transport behavior of FQs under Type-III MC configuration at $B/B^\Phi = 0.5$. The representations are the same as those in Video 1.

**Video 3 | DC Transport Property under Type-III MC Configuration at $B/B^\Phi = 1.0$**

Video showing asymmetric DC transport behavior of FQs under Type-III MC configuration at $B/B^\Phi = 1.0$. Combined with Video 2, it shows reversible DC transport behavior of FQs at different flux quanta density.